\documentclass[twocolumn,twocolappendix,times]{aastex631}

\usepackage{savesym}  
\savesymbol{splitbox}
\usepackage[export]{adjustbox}
\restoresymbol{TXF}{splitbox}
\usepackage{multirow}   
\usepackage{amsmath}
\usepackage{CJK}

\newcommand{\fracA}{{\rm frac_{AGN}}}
\newcommand{\nmad}{\sigma_{\rm NMAD}}
\newcommand{\fracOut}{{\rm frac_{out}}}

\newcommand{\fst}[1]{#1}

\shorttitle{MIRI AGN}
\shortauthors{Yang et al.}

\graphicspath{{./}{figures/}}

\begin{document}
\begin{CJK*}{UTF8}{gbsn}

\title{CEERS Key Paper VI: JWST/MIRI Uncovers a Large Population of Obscured AGN at High Redshifts}
  
\correspondingauthor{Guang Yang}
\email{gyang206265@gmail.com}

\author[0000-0001-8835-7722]{Yang, G. (杨光)}
\affiliation{Kapteyn Astronomical Institute, University of Groningen, P.O. Box 800, 9700 AV Groningen, The Netherlands}
\affiliation{SRON Netherlands Institute for Space Research, Postbus 800, 9700 AV Groningen, The Netherlands}

\author{Caputi, K. I.}
\affiliation{Kapteyn Astronomical Institute, University of Groningen, P.O. Box 800, 9700 AV Groningen, The Netherlands}
\affiliation{Cosmic Dawn Center (DAWN), Copenhagen, Denmark}

\author[0000-0001-7503-8482]{Papovich, C.}
\affiliation{Department of Physics and Astronomy, Texas A\&M University, College
Station, TX, 77843-4242 USA}
\affiliation{George P.\ and Cynthia Woods Mitchell Institute for
 Fundamental Physics and Astronomy, Texas A\&M University, College
 Station, TX, 77843-4242 USA}

 \author[0000-0002-7959-8783]{Arrabal Haro, P.}
\affiliation{NSF's National Optical-Infrared Astronomy Research Laboratory, 950 N. Cherry Ave., Tucson, AZ 85719, USA}

 \author[0000-0002-9921-9218]{Bagley, M. B.}
\affiliation{Department of Astronomy, The University of Texas at Austin, Austin, TX, USA}

\author[0000-0002-2517-6446]{Behroozi, P.}
\affiliation{Department of Astronomy and Steward Observatory, University of Arizona, Tucson, AZ 85721, USA}
\affiliation{Division of Science, National Astronomical Observatory of Japan, 2-21-1 Osawa, Mitaka, Tokyo 181-8588, Japan}

\author[0000-0002-5564-9873]{Bell, E. F.}
\affiliation{Department of Astronomy, University of Michigan, 1085 S. University Ave, Ann Arbor, MI 48109-1107, USA}

\author[0000-0003-0492-4924]{Bisigello, L.}
\affiliation{Dipartimento di Fisica e Astronomia "G.Galilei", Universit\'a di Padova, Via Marzolo 8, I-35131 Padova, Italy}
\affiliation{INAF--Osservatorio Astronomico di Padova, Vicolo dell'Osservatorio 5, I-35122, Padova, Italy}

\author[0000-0003-3441-903X]{Buat, V.}
\affiliation{Aix Marseille Univ, CNRS, CNES, LAM Marseille, France}

\author[0000-0002-4193-2539]{Burgarella, D.}
\affiliation{Aix Marseille Univ, CNRS, CNES, LAM Marseille, France}

\author[0000-0001-8551-071X]{Cheng, Y.}
\affiliation{University of Massachusetts Amherst, 710 North Pleasant Street, Amherst, MA 01003-9305, USA}

\author[0000-0001-7151-009X]{Cleri, N. J.}
\affiliation{Department of Physics and Astronomy, Texas A\&M University, College Station, TX, 77843-4242 USA}
\affiliation{George P.\ and Cynthia Woods Mitchell Institute for Fundamental Physics and Astronomy, Texas A\&M University, College Station, TX, 77843-4242 USA}

\author[0000-0003-2842-9434]{Dav\'e, R.}
\affiliation{Institute for Astronomy, University of Edinburgh, Blackford Hill, Edinburgh, EH9 3HJ UK}
\affiliation{Department of Physics and Astronomy, University of the Western Cape, Robert Sobukwe Rd, Bellville, Cape Town 7535, South Africa}

\author[0000-0001-5414-5131]{Dickinson, M.}
\affiliation{NSF's National Optical-Infrared Astronomy Research Laboratory, 950 N. Cherry Ave., Tucson, AZ 85719, USA}

\author[0000-0002-7631-647X]{Elbaz, D.}
\affiliation{Universit{\'e} Paris-Saclay, Universit{\'e} Paris Cit{\'e}, CEA, CNRS, AIM, 91191, Gif-sur-Yvette, France}

\author[0000-0001-7113-2738]{Ferguson, H. C.}
\affiliation{Space Telescope Science Institute, Baltimore, MD, USA}

\author[0000-0001-8519-1130]{Finkelstein, S. L.}
\affiliation{Department of Astronomy, The University of Texas at Austin, Austin, TX, USA}

\author[0000-0001-9440-8872]{Grogin, N. A.}
\affiliation{Space Telescope Science Institute, 3700 San Martin Dr., Baltimore, MD 21218, USA}

\author[0000-0001-6145-5090]{Hathi, N. P.}
\affiliation{Space Telescope Science Institute, Baltimore, MD, USA}

\author[0000-0002-3301-3321]{Hirschmann, M.}
\affiliation{Institute of Physics, Laboratory of Galaxy Evolution, Ecole Polytechnique F\'ed\'erale de Lausanne (EPFL), Observatoire de Sauverny, 1290 Versoix, Switzerland}

\author[0000-0002-4884-6756]{Holwerda, B. W.}
\affil{Physics \& Astronomy Department, University of Louisville, 40292 KY, Louisville, USA}

\author[0000-0002-1416-8483]{Huertas-Company, M.}
\affil{Instituto de Astrof\'isica de Canarias, La Laguna, Tenerife, Spain}
\affil{Universidad de la Laguna, La Laguna, Tenerife, Spain}
\affil{Universit\'e Paris-Cit\'e, LERMA - Observatoire de Paris, PSL, Paris, France}

\author[0000-0001-6251-4988]{Hutchison, T. A.}
\altaffiliation{NASA Postdoctoral Fellow}
\affiliation{Astrophysics Science Division, NASA Goddard Space Flight Center, 8800 Greenbelt Rd, Greenbelt, MD 20771, USA}

\author[0000-0001-8386-3546]{Iani, E.}
\affiliation{Kapteyn Astronomical Institute, University of Groningen, P.O. Box 800, 9700 AV Groningen, The Netherlands}

\author[0000-0001-9187-3605]{Kartaltepe, J. S.}
\affiliation{Laboratory for Multiwavelength Astrophysics, School of Physics and Astronomy, Rochester Institute of Technology, 84 Lomb Memorial Drive, Rochester, NY 14623, USA}

\author[0000-0002-5537-8110]{Kirkpatrick, A.}
\affiliation{Department of Physics and Astronomy, University of Kansas, Lawrence, KS 66045, USA}

\author[0000-0002-8360-3880]{Kocevski, D. D.}
\affiliation{Department of Physics and Astronomy, Colby College, Waterville, ME 04901, USA}

\author[0000-0002-6610-2048]{Koekemoer, A. M.}
\affiliation{Space Telescope Science Institute, 3700 San Martin Dr., Baltimore, MD 21218, USA}

\author[0000-0002-5588-9156]{Kokorev, V.}
\affiliation{Kapteyn Astronomical Institute, University of Groningen, P.O. Box 800, 9700 AV Groningen, The Netherlands}

\author[0000-0003-2366-8858]{Larson, R. L.}
\affiliation{NSF Graduate Fellow}
\affiliation{The University of Texas at Austin, Department of Astronomy, Austin, TX, United States}

\author[0000-0003-1581-7825]{Lucas, R. A.}
\affiliation{Space Telescope Science Institute, 3700 San Martin Drive, Baltimore, MD 21218, USA}

\author[0000-0003-4528-5639]{P\'erez-Gonz\'alez, P. G.}
\affiliation{Centro de Astrobiolog\'{\i}a (CAB), CSIC-INTA, Ctra. de Ajalvir km 4, Torrej\'on de Ardoz, E-28850, Madrid, Spain}

\author{Rinaldi, P.}
\affiliation{Kapteyn Astronomical Institute, University of Groningen, P.O. Box 800, 9700 AV Groningen, The Netherlands}

\author[0000-0001-9495-7759]{Shen, L.}
\affiliation{Department of Physics and Astronomy, Texas A\&M University, College
Station, TX, 77843-4242 USA}
\affiliation{George P.\ and Cynthia Woods Mitchell Institute for
 Fundamental Physics and Astronomy, Texas A\&M University, College
 Station, TX, 77843-4242 USA}

 \author[0000-0002-1410-0470]{Trump, J. R.}
\affiliation{Department of Physics, 196 Auditorium Road, Unit 3046, University of Connecticut, Storrs, CT 06269, USA}

\author[0000-0002-6219-5558]{de la Vega, A.}
\affiliation{Department of Physics and Astronomy, University of California, 900 University Ave, Riverside, CA 92521, USA}

\author[0000-0003-3466-035X]{Yung, L. Y. A.}
\affiliation{Astrophysics Science Division, NASA Goddard Space Flight Center, 8800 Greenbelt Rd, Greenbelt, MD 20771, USA}

\author[0000-0002-7051-1100]{Zavala, J, A.}
\affiliation{National Astronomical Observatory of Japan, 2-21-1 Osawa, Mitaka, Tokyo 181-8588, Japan}

\begin{abstract}

Mid-infrared observations are powerful in identifying heavily obscured Active Galactic Nuclei (AGN) which have weak emission in other wavelengths. 
\fst{Data from the Mid-Infrared Instrument (MIRI) onboard \textit{JWST} provides an excellent opportunity} to perform such studies.
We take advantage of the MIRI imaging data from the Cosmic Evolution Early Release Science Survey (CEERS) to investigate the AGN population in the distant universe. 
We estimate the source properties of MIRI-selected objects by utilizing spectral energy distribution (SED) modelling, and classify them into star-forming galaxies (SF), SF-AGN mixed objects, and AGN. The source numbers of these types are 433, 102, and 25, respectively, from 4 MIRI pointings covering $\sim 9$~arcmin$^2$. 
The sample spans a redshift range of $\approx 0$--5. 
We derive the median SEDs for all three source types, respectively, and publicly release them.
The median MIRI SED of AGN is similar to the typical SEDs of hot dust-obscured galaxies and Seyfert~2s, for which the mid-IR SEDs are \fst{dominated by emission} from AGN-heated hot dust.
Based on our SED-fit results, we estimate the black-hole accretion density (BHAD; i.e., total BH growth rate per comoving volume) as a function of redshift. 
At $z<3$, the resulting BHAD agrees with the X-ray measurements in general.
At $z>3$, we identify a total of 27 AGN and SF-AGN mixed objects, leading to that our high-$z$ BHAD is substantially higher than the X-ray results ($\sim 0.5$~dex at $z \approx 3$--5).
This difference indicates MIRI can identify a large population of heavily obscured AGN missed by X-ray surveys at high redshifts. 

\end{abstract}

\section{Introduction} 
\label{sec:intro}
Active Galactic Nuclei (AGN) are powered by growing supermassive black holes (BHs) typically harboured at the centers of galaxies.
AGN luminosities can be used to estimate BH growth rates.
From studying the AGN demography at different redshifts, the evolution history of the BH population can be inferred \fst{\citep[e.g.,][]{merloni08, yang18, aird19, aird22, ananna19}}. 
To depict an unbiased picture of the BH cosmic evolution, it is critical to conduct a complete census of the AGN population minimizing selection biases. 

Multiwavelength techniques of AGN selection have been intensively developed in the past $\sim 60$ years since their discovery (e.g., \fst{\citealt{padovani17}}; \citealt{hickox18,lyu22}).
Among these techniques, \fst{optical AGN selection is often the most straightforward and can be performed with ground-based telescopes}. 
Color-color diagrams based on broad-band photometry are often employed to select AGN candidates \citep[e.g.,][]{richards02}.
\fst{Follow-up spectroscopic analyses identifying broad lines or using line ratios \citep[e.g.,][]{kauffmann03} can verify the AGN nature.
However, optical color-color selections are biased toward selecting} relatively unobscured type~1 AGN. 
Compared to optical selections, X-ray AGN selections are more complete in general.
Strong X-ray emission is ubiquitous among the AGN population, and X-ray photons, especially hard X-rays, have excellent penetrating power. 
These advantages allow X-ray methods to select not only unobscured AGN but also obscured ones \citep[e.g.,][]{brandt15, brandt21}. 
However, even the hardest X-ray photons with $E>10$~keV can be Compton scattered and absorbed when the obscuration is heavy with column density $N_{\rm H} \gtrsim 10^{24}\rm\ cm^{-2}$.
Indeed, evidence suggests that X-ray surveys might miss many heavily obscured AGN, especially at high redshifts ($z\gtrsim 3$) where the AGN obscuration is generally strong, because gas tends to be more abundant and compact toward high redshift \citep[e.g.,][]{yang21b, gilli22}. 

Infrared (IR) observations have the \fst{distinct} advantage of selecting obscured AGN. 
Obscured AGN are often dust-rich, and these dust particles are heated to $\sim 100$--1000~K by the emission from the central engine of the accretion disk.  
The dust re-emits the ``waste heat'' primarily in the mid-IR wavelengths of $\sim 3$--30~$\mu$m. 
By detecting this mid-IR radiation, AGN can be identified. 
Color-color methods based on \textit{Spitzer}/Infrared Array Camera (IRAC) have been developed and widely applied to IRAC surveys \citep[e.g.,][]{stern05, lacy07, donley12}. 
Studies based on \textit{Spitzer} data have also worked on the selection of AGN based on the presence of a power law in the IRAC bands
\citep[e.g.,][]{alonso_herrero06, caputi13}.
These techniques have successfully identified numerous obscured AGN, some of which are missed by even the deepest X-ray surveys \citep[e.g.,][]{lyu22}. 
However, due to the limitations of IRAC sensitivity and wavelength coverage (below 8~$\mu$m), the techniques are ineffective in identifying \fst{extremely} faint AGN, especially at high redshifts, where AGN radiation is redshifted beyond the IRAC wavelength coverage \citep{mendez13, lyu22b}. 
The Multiband Imaging Photometer for \textit{Spitzer} (MIPS) and \textit{Herschel} cover longer wavelengths, but they are not \fst{comparably} suitable for distant-AGN searches due to their lower sensitivities compared to IRAC.  

The Mid-Infrared Instrument (MIRI) onboard the recently launched \textit{James Webb Space Telescope} (\textit{JWST}) has the potential to become a game changer in the detection and selection of AGN especially at high redshifts. 
MIRI has continuous photometry coverage of $\sim 5$--25~$\mu$m, reaching an unprecedented sensitivity level about $10$--100 times deeper than \textit{Spitzer}. 
Besides, thanks to the large aperture size of \textit{JWST}, the angular resolution of MIRI reaches a full-width half maximum (FWHM) of point spread functions (PSFs) being $\sim 0.2$--0.8$''$, about 8 times better than \textit{Spitzer}/IRAC at similar wavelengths.
Indeed, simulations  suggest that MIRI is able to capture AGN activity down to a faint level of, e.g., Eddington ratio around $0.01$ at $z \sim 1$--2 \citep[e.g.,][]{kirkpatrick17,yang21}. 

In this paper, we take advantage of the MIRI imaging data from the Cosmic Evolution Early Release Science Survey (CEERS; \citealt{finkelstein17}), which targets the Extended Groth Strip (EGS) field.
We also compile the multiwavelength data available in EGS and perform AGN versus star-forming (SF) classifications with spectral energy distribution (SED) modeling. 
We produce median SEDs for different source types, respectively, and publicly release these empirical SEDs to facilitate future science. 
Finally, we investigate the BH accretion density (BHAD) as a function of redshift, based on the MIRI-identified AGN. 

The structure of this paper is as follows.
In \S\ref{sec:analyis}, we describe and analyze the MIRI imaging data. 
We also perform AGN selections based on the MIRI and other multiwavelength photometric data. 
In \S\ref{sec:discuss}, we discuss the physical implications based on our analysis results. 
We summarize our results and discuss future prospects in \S\ref{sec:sum}.

Throughout this paper, we assume a cosmology with $H_0=70$~km~s$^{-1}$~Mpc$^{-1}$, $\Omega_M=0.3$, and $\Omega_{\Lambda}=0.7$.
We adopt a Chabrier initial mass function (IMF; \hbox{\citealt{chabrier03}}).
Quoted uncertainties are at the $1\sigma$\ (68\%) confidence level.
All magnitudes are in AB units \citep{oke83}, 
where $m_\mathrm{AB} = -48.6 - 2.5 \log(f_\nu)$ for $f_\nu$ in units of 
erg~s$^{-1}$~cm$^{-2}$ Hz$^{-1}$.   

\section{Data and analysis}
\label{sec:analyis}

\subsection{MIRI and other multi-wavelength photometry}
\label{sec:photo}
The CEERS survey has 8 MIRI pointings: 4 ``blue'' and 4 ``red''. 
Each blue pointing has two bands of F560W and F770W. 
Two of the red pointings have six bands from F770W to F2100W, and the other two have four bands from F1000W to F1800W.  
In this paper, we focus on the 4 red pointings, as the blue pointings lack the long mid-IR wavelength coverage for AGN identification (\S\ref{sec:intro}).
The MIRI photometry is extracted with \textsc{t-phot} \citep{merlin15}, using \textit{Hubble Space Telescope (HST)}/F160W as the high-resolution prior. 
The details of the MIRI observations, data reduction, photometry extraction, and quality assessment will be presented in a dedicated paper (Yang et al. in prep.), and some methods are discussed in \cite{yang21} and \cite{papovich23}.

We also compile other multiwavelength photometric data, specifically the EGS catalog by \cite{stefanon17} from the Cosmic Assembly Near-infrared Deep Extragalactic Legacy Survey \citep[CANDELS:][]{grogin11,koekemoer11}.
These data include 17 broad-band photometry from the $U$ band to the IRAC $8\ \mu$m.

Because this paper is focused on the MIRI properties of galaxies, we restrict our sample to objects with at least two MIRI bands with detections of signal-to-noise ($\rm S/N$) above 3.  
This leads to a parent sample of 560 sources.
This criterion guarantees that we have at least one robust MIRI color for each source.  
In general, shorter-wavelength MIRI bands have higher detection rates thanks to their relatively deeper sensitivity. 
For example, the $\rm S/N>3$ detection rates of F770W and F2100W are 97\% and 68\%, respectively, among our sources. 

\subsection{SED modelling setup}
\label{sec:sed}
We employ \textsc{cigale} v2022.1 \citep{boquien19, yang20, yang22} to perform SED modeling based on the broad-band photometry described in \S\ref{sec:photo}.  
The adopted \textsc{cigale} parameters are summarized in Table~\ref{tab:cig}.

For star formation history (SFH),
we use a standard delayed-$\tau$ module (\texttt{sfhdelayed} in \textsc{cigale}). 
We allow the $e$-folding time and stellar age varying from 0.5--5~Gyr and 1--5~Gyr, respectively. 
We employ \cite{bruzual03} (\texttt{bc03}) for the simple stellar population (SSP) module, assuming a \cite{chabrier03} initial mass function (IMF) with a solar metallicity ($Z=0.02$).
We also include the \texttt{nebular} module \citep{villa-velez21} for emission from the HII regions. 
For the dust attenuation, we adopt \texttt{dustatt\_modified\_starburst} in \textsc{cigale}, which is primarily based on \cite{calzetti00} extending to short wavelengths (91.2--150~nm) with \cite{leitherer02}. 
The allowed range of  color excess is $E(B-V)=0$--1 (see Table~\ref{tab:cig}).

For galactic dust emission, we use the \texttt{dl2014} module \citep{draine14}.
The dust luminosity numerically equals the attenuated luminosity of starlight, because \textsc{cigale} strictly follows energy conservation. 
The \citep{draine14} recipe models the dust emission with two components, a diffused emission and a photodissociation region (PDR) emission associated with star formation. 
The two components share the same mass fraction of polycyclic aromatic hydrocarbon (PAH) compared to total dust, and we allow three different values of it (see Table~\ref{tab:cig}).
The major difference between the two components is the radiation field. 
The diffuse dust is radiated by the minimum radiation parameter ($U_{\rm min}$) set by the user, and the PDR's radiation field ranges from $U_{\rm min}$ to a maximum fixed value of $U_{\rm max}=10^7$.
We allow $U_{\rm min}$ varying from 0.1, 1.0, 10, 50.
The relative strength between the two components is controlled by the $\gamma$ parameter (the fraction of PDR emission) and we allow it \fst{to vary in the range} 0.01--0.9.

For the AGN component, we adopt the \texttt{skirtor2016} module based on a clumpy torus model from \cite{stalevski12, stalevski16}.
The relative strength between the AGN and galaxy components are set by the $\fracA$ parameter, i.e., AGN fractional luminosity compared to the total over a user-defined rest-frame wavelength range ($\lambda_{\rm AGN}$). 
In this work, we allow $\fracA$ \fst{to vary} from 0 to 0.99, and set $\lambda_{\rm AGN}$ to a mid-IR wavelength range of 3--30~$\mu$m which contains the bulk of AGN dust emission (Table~\ref{tab:cig}). 
We allow all available values \fst{of the} 9.7~$\mu$m optical depth (i.e., 3, 5, 7, 9, and 11) in \texttt{skirtor2016}. 
The parameter of AGN viewing angle determines the AGN type. 
Following \cite{yang21}, we only allow type~2 AGN, because our focus in this work is to find the obscured AGN untraceable in UV/optical.   
Including type~1 models would cause a significant model degeneracy versus the stellar emission in UV/optical \citep[e.g.,][]{ni21, zou22}.
Also, type~1 AGN are rare especially for small-area surveys like our case, and we expect only $\lesssim 1$ type~1 AGN using the empirical number density in \cite{yang18}.
We fix the viewing angle at 70$^\circ$, a typical value for type~2 AGN \citep{yang20}, because the fit results are generally insensitive to different type~2 viewing angles \citep[e.g.,][]{ramos22}.
\fst{Here, by using \texttt{skirtor2016} we assume AGN obscuration is mainly caused by dust on the torus scale ($\sim$~pc) rather than on the galactic scale ($\gtrsim$~100~pc). 
This is because galactic-scale dust is likely heated to low temperature ($\lesssim 100$~K), and consequently the reprocessed emission concentrates on far-IR, beyond the MIRI wavelength coverage. 
We discuss the effect of possible galactic-scale obscuration in \S\ref{sec:caveat}.
}

Since most (95\%) of our sources do not have spectroscopic redshifts (spec-$z$) available, we allow \textsc{cigale} to fit a redshift for each source.\footnote{For consistency across the sample, we also fit the redshift for spec-$z$ sources.
We note that fixing their redshifts at spec-$z$ does not affect our main conclusions.} 
We set the redshift grid from 0.01 to 8.0 (Table~\ref{tab:cig}) with 50 steps evenly spaced in terms of $\log(1+z)$.
We note that changing the maximum allowed redshift ($z=8$) to higher values \fst{does} not affect the fit results.
We evaluate the best-fit photometric redshift (photo-$z$) quality using the available spec-$z$ \fst{measurements} in \S\ref{sec:sed_res}.

\begin{table*}
\centering
\caption{\textsc{cigale} model parameters}
\label{tab:cig}
\begin{tabular}{llll} \hline\hline
Module & Parameter & Symbol & Values \\
\hline
\multirow{2}{*}{\shortstack[l]{Star formation history\\
                               \texttt{sfhdelayed} }}
    & Stellar e-folding time & $\tau_{\rm star}$ & 0.5, 1, 2, 3, 4, 5 Gyr\\
    & Stellar age & $t_{\rm star}$  
            & \fst{0.1, 0.2, 0.5,} 1, 2, 3, 4, 5 Gyr\\ 
\hline
\multirow{2}{*}{\shortstack[l]{Simple stellar population\\ 
    \texttt{bc03} }}
    & Initial mass function & $-$ & \cite{chabrier03} \\
    & Metallicity & $Z$ & 0.02 \\
\hline
\multirow{2}{*}{\shortstack[l]{Nebular emission\\
                               \texttt{nebular} }} 
    & Ionization parameter & $\log U$ & $-2.0$\\
    & Gas metallicity & $Z_{\rm gas}$ & 0.02 \\
\hline
\multirow{3}{*}{\shortstack[l]{Dust attenuation \\ 
                \texttt{dustatt\_modified\_starburst} }}
    & \multirow{2}{*}{\shortstack[l]{Color excess of nebular lines}} & \multirow{2}{*}{\shortstack[l]{$E(B-V)_{\rm line}$}} & \multirow{2}{*}{\shortstack[l]{0,0.02,0.05,0.1,0.2,\\
                              0.3,0.4,0.5,0.7,0.9,1.0}} \\\\
    & ratio between line and continuum $E(B-V)$ & $\frac{E(B-V)_{\rm line}}{E(B-V)_{\rm cont}}$ & 1 \\
\hline
\multirow{4}{*}{\shortstack[l]{Galactic dust emission \\ \texttt{dl2014} }}
    & PAH mass fraction & $q_{\rm PAH}$ & 0.47, 2.5, 7.32 \\
    & Minimum radiation field & $U_{\rm min}$ & 0.1, 1.0, 10, 50 \\
    & \multirow{2}{*}{\shortstack[l]{Fraction of PDR emission}} & \multirow{2}{*}{\shortstack[l]{$\gamma$}} & \multirow{2}{*}{\shortstack[l]{0.01, 0.02, 0.05, \\
                                            0.1, 0.2, 0.5, 0.9}} \\\\
\hline
\multirow{5}{*}{\shortstack[l]{AGN (UV-to-IR) emission \\ \texttt{skirtor2016} }}
    & Average edge-on optical depth at $9.7 \mu$m & $\tau_{9.7}$ & 3,5,7,9,11 \\
    & Viewing angle & $\theta_{\rm AGN}$ & 70$^\circ$ \\
    & \multirow{2}{*}{\shortstack[l]{AGN contribution to IR luminosity}} & \multirow{2}{*}{\shortstack[l]{$\fracA$}} & \multirow{2}{*}{\shortstack[l]{0, 0.01, 0.03, 0.05, \\ 
    0.1--0.9 (step 0.1), 0.99  }}\\\\
    & Wavelength range where $\fracA$ is defined & $\lambda_{\rm AGN}$ & 3--30 $\mu$m \\
\hline
    \multirow{2}{*}{\shortstack[l]{Redshift$+$IGM \\ \texttt{redshifting} }} 
    &  \multirow{2}{*}{\shortstack[l]{ Source redshift }} & \multirow{2}{*}{\shortstack[l]{ $z$ }} & \multirow{2}{*}{\shortstack[l]{0.01--8.0} (50 steps)} \\\\
\hline
\end{tabular}
\begin{flushleft}
{\sc Note.} --- For parameters not listed here, default values are being adopted.
\end{flushleft}
\end{table*}

\subsection{CIGALE results}
\label{sec:sed_res}
\begin{figure*}    
    \centering
    \includegraphics[width=2\columnwidth]{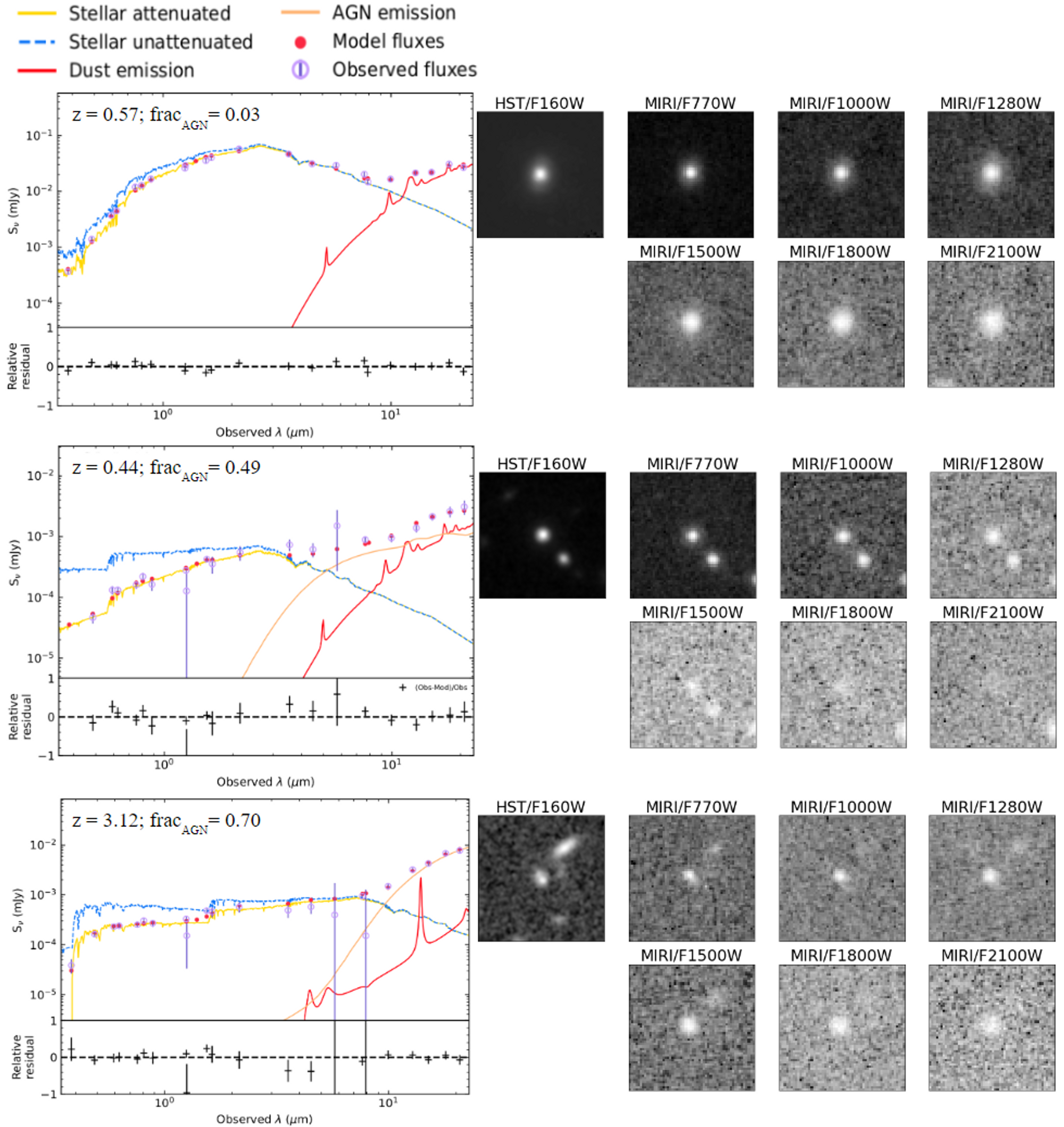}
    \caption{\textit{Left}: SED-fit examples from \textsc{cigale}.
    The object in the top/middle/bottom panel is classified as SF/mixed/AGN.
    Within each panel, the top sub-panel displays the data points and the best-fit SED models (i.e., flux density versus observed wavelength). 
    The purple and red data points indicate the observed and model flux densities, respectively. 
    The black curves represent the total model SEDs. 
    The red and orange curves indicate the galactic dust and AGN components, respectively. 
    The blue/yellow curves represent unattenuated/attenuated stellar.
    The fitted redshift and $\fracA$ are labeled.
    The bottom sub-panel display the relative residual \fst{in the flux}, i.e., (observed$-$model)/observed.
    \textit{Right}: The corresponding $5''\times 5''$ cutouts from \textit{HST}/F160W and MIRI. 
    }
    \label{fig:sed}
\end{figure*}

We run \textsc{cigale} with the settings \fst{described} in \S\ref{sec:sed}.
We adopt the Bayesian output instead of the best-fit output.
The former is probability-weighted values considering all models available, while the latter is from the minimum-$\chi^2$ model alone.
Therefore, the Bayesian output is generally more robust than the best-fit output.
Fig.~\ref{fig:sed} displays three SED-fit examples from our run and the corresponding image cutouts.

\begin{figure}
    \centering
    \includegraphics[width=\columnwidth]{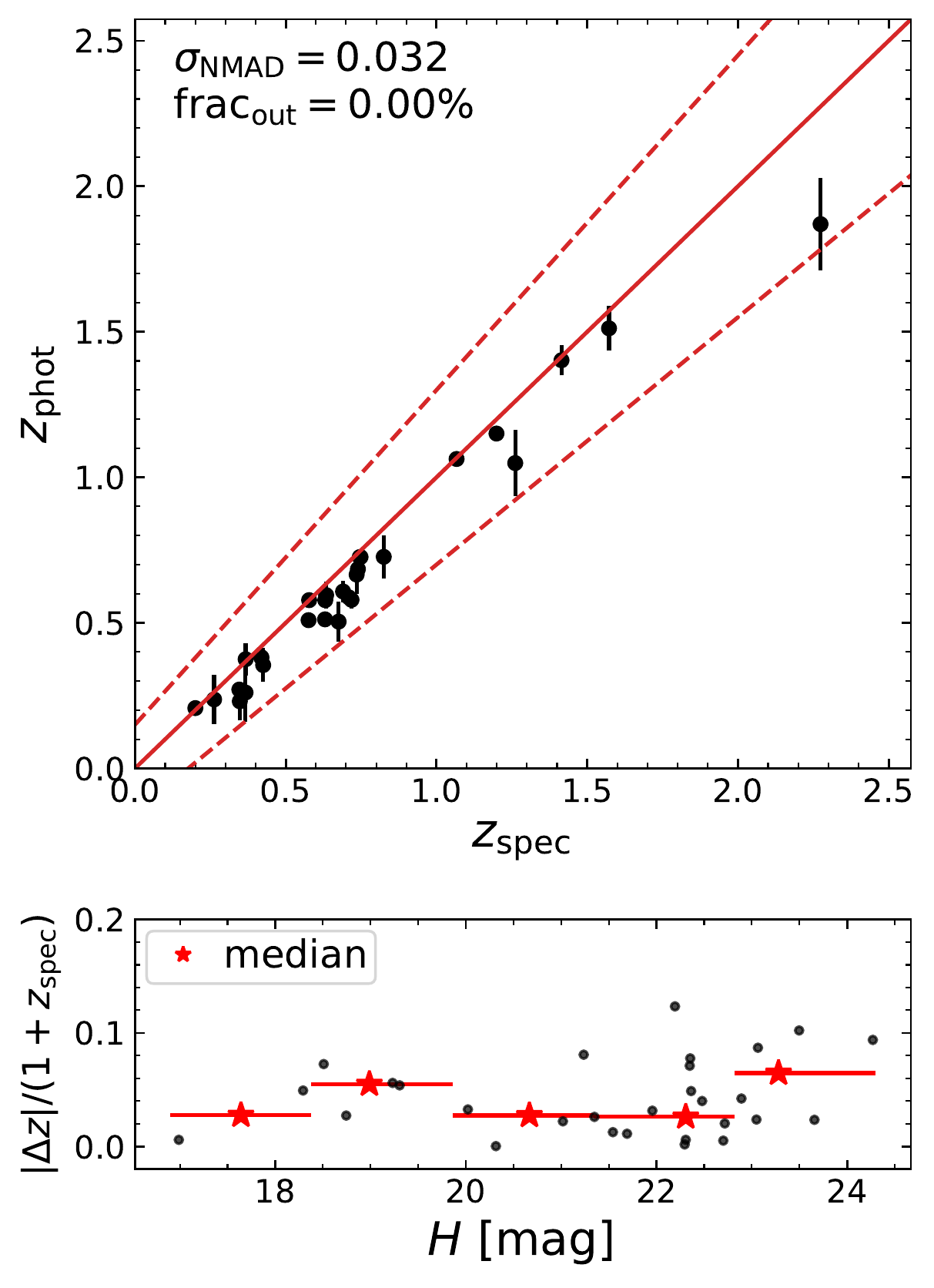}
    \caption{\textit{Top}: Comparison between our photometric redshifts (from \textsc{cigale} fits) versus spectroscopic redshifts (when available).
    The red solid line indicates the 1:1 relation, while the red dashed lines indicate a 15\% uncertainty (beyond which a source would be considered as a photo-$z$ outlier). 
    The photo-$z$ quality, quantified by $\nmad$ and $\fracOut$, is marked.
    We note that the outlier fraction is zero. 
    \fst{\textit{Bottom}: Relative difference between photo-$z$ and spec-$z$ as a function of $H$-band magnitude. 
    Each red star represents the median value for each magnitude bin, while the horizontal error bar indicates the bin coverage. 
    It appears the photo-$z$ quality does not significantly depend on the magnitude. 
    }
    }
    \label{fig:z_vs_z}
\end{figure}

A small subsample of our sources (29 out of 560) have secure spec-$z$ measurements available from the CANDEL-EGS catalog \citep{stefanon17}.
We compare these spec-$z$ values versus our photo-$z$ estimation in Fig.~\ref{fig:z_vs_z} (top).
We assess the photo-$z$ quality by two parameters, i.e., normalized median absolute
deviation ($\nmad$) and outlier fraction ($\fracOut$).
The former is defined as $\nmad=1.48 \times {\rm median}\{|\Delta z - {\rm median}(\Delta z)|/(1+z\_{\rm spec})\}$, where $\Delta z$ is the difference between photo-$z$ and spec-$z$;
the latter is defined as the fraction of outliers [defined as $|\Delta z|/(1+z\_{\rm spec}) > 0.15$] among all spec-$z$ sources.
We obtain satisfactory photo-$z$ quality of $\nmad=0.032$ and $\fracOut=0.00\%$ for our spec-$z$ subsample.
We attribute the remarkable zero outlier fraction to the MIRI coverage of the mid-IR PAH features and CANDELS UV-to-near-IR coverage of Lyman and/or Balmer breaks. 
Indeed, previous simulations have indicated that MIRI photometry could potentially reduce the number of photo-$z$ outliers significantly, thanks to the coverage of the PAH features as a redshift indicator \citep[e.g.,][]{bisigello16, yang21}.
From Fig.~\ref{fig:z_vs_z}, the measured photo-$z$ tends to be systematically lower than the spec-$z$ (median offset $\sim 3\%$).
The exact cause is not clear, but this minor systematic \fst{does} not affect our main results.  
\fst{To assess the dependence of photo-$z$ quality on source brightness, we also plot the photo-$z$ error (compared to spec-$z$) vs.\ the $H$-band magnitude.
It appears that there is no obvious trend between the error and the magnitude. 
}

\begin{figure}
    \centering
    \includegraphics[width=\columnwidth]{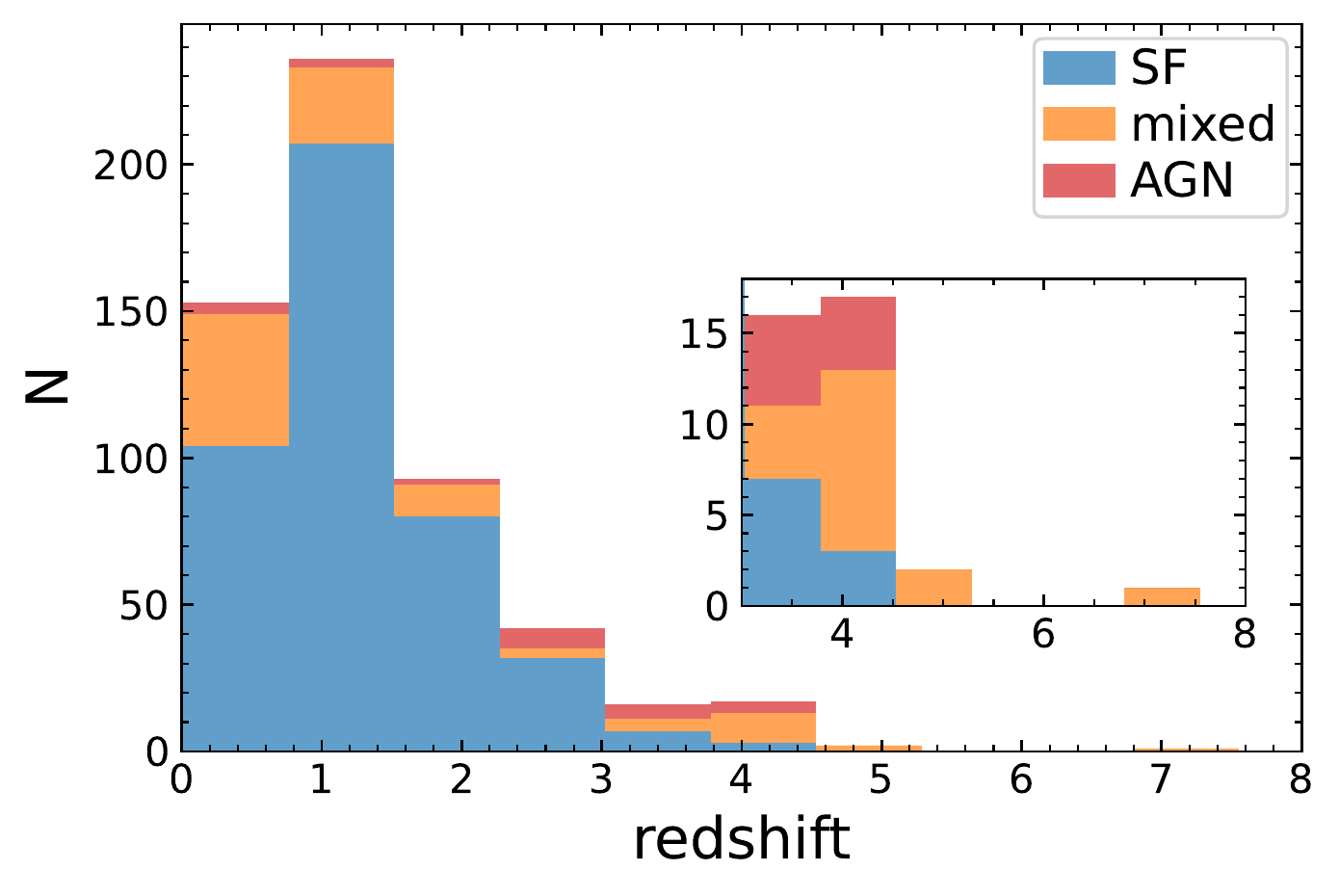}
    \caption{Photometric redshift distribution of our MIRI sources. 
    The blue, orange, and red bars represent SF, mixed, and AGN types, respectively. 
    The inset panel zooms in on the $z\gtrsim 3$ part of the histogram. 
    }
    \label{fig:z_hist}
\end{figure}

We classify \fst{each source as a star-forming galaxy, a star forming-AGN mixed system, or a pure AGN} (SF, mixed, and AGN, hereafter), based on the best-fit $\fracA$ parameter (see \S\ref{sec:sed} for its definition). 
For an object with $\fracA \leq 0.1$, $0.1< \fracA < 0.5$, and $\fracA \geq 0.5$, we classify it as SF, mixed, and AGN, respectively.
From these criteria, 433, 102, and 25 objects are classified as SF, mixed, and AGN, respectively. 
We note that slightly adjusting these empirical classification criteria \fst{does} not change our main results qualitatively.
Fig.~\ref{fig:z_hist} displays the photo-$z$ distributions for different object types.


The CEERS field has X-ray coverage from \textit{Chandra} \citep{nandra15}. 
Among our MIRI sources, there are 8 X-ray detected AGN (selected as $L_X > 10^{42.5}$~erg~s$^{-1}$; e.g., \citealt{yang18}), and 1, 4, and 3 of these eight X-ray AGN are classified as SF, mixed, and AGN, respectively, by our SED method above. 
Therefore, our classifications successfully retrieved $7/8=87.5\%$ of the X-ray AGN. 
The only X-ray AGN misclassified as SF has $\fracA=0.08 \pm 0.06$, marginally below our threshold for the mixed type ($\fracA=0.1$).

\section{Discussion}
\label{sec:discuss}
In this section, we present the mid-IR SEDs for SF, mixed, and AGN types (\S\ref{sec:comp_sed}) and discuss the cosmic evolution history of supermassive black holes (\S\ref{sec:bhad}).
We give some caveats for our results in \S\ref{sec:caveat}.

\subsection{Median SEDs}
\label{sec:comp_sed}
In \S\ref{sec:sed_res}, we have classified each source as SF, mixed, or AGN based on the SED-fit result.
We now study the ensemble SED features for each source type from their median SEDs in the rest frame. 

\begin{figure*}
    \centering    \includegraphics[width=2\columnwidth]{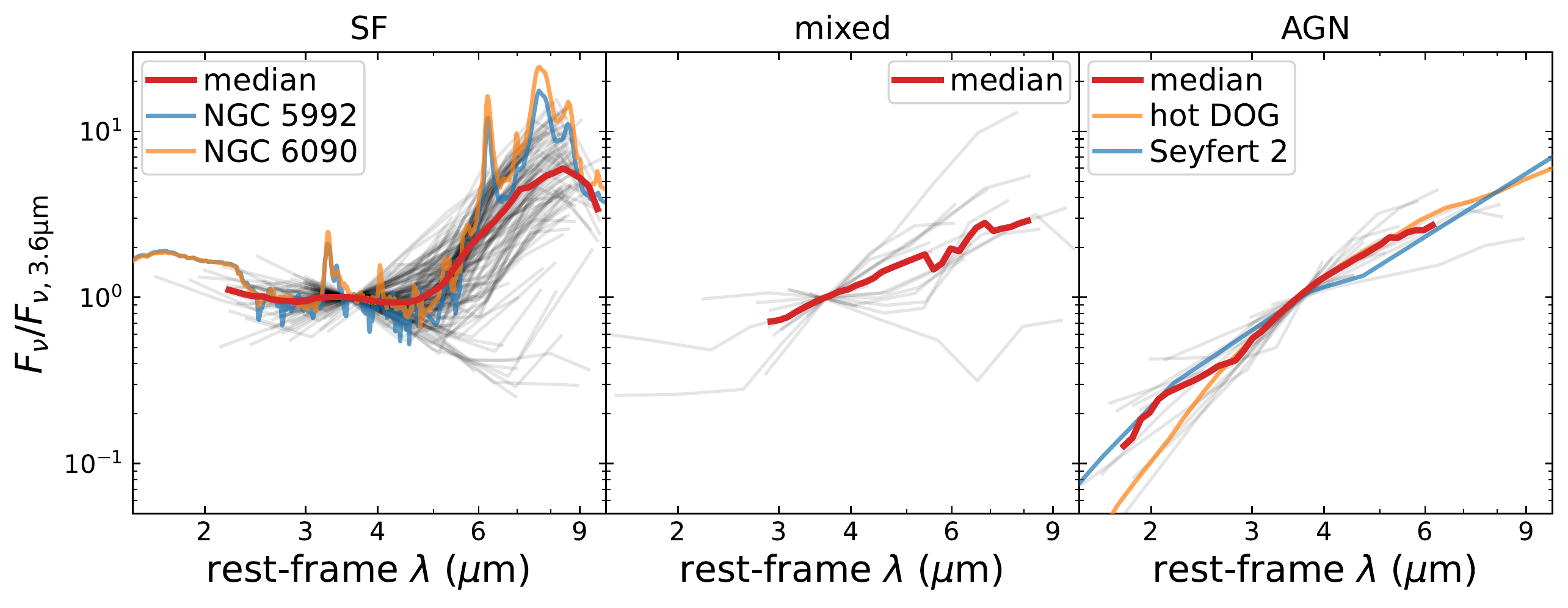}
    \caption{Median rest-frame SEDs of SF (left), mixed (middle), and AGN (right) types.
    The red curves indicate the median SEDs.
    The grey curves indicate the MIRI SEDs (only $S/N>3$ MIRI points are being used here for high quality) for individual sources. 
    All these individual SEDs are normalized at rest-frame 3.6~$\mu$m. 
    The median SEDs are estimated from these grey curves. 
    On the left panel, we also display two local SF galaxy SEDs, NGC~5992 and NGC~6090 from \cite{brown14}.
    Our median SF SED is similar to the SEDs of these two local SF galaxies.
    On the right panel, we also display the median SED for hot dust-obscured galaxies (hot DOG; \citealt{fan16}) and the average SED for the nuclear regions of Seyfert~2 galaxies \citep{videla13}.
    These typical SEDs are similar to our MIRI-selected AGN median SED.
    }
    \label{fig:composite_sed}
\end{figure*}

To estimate the median SED,
we first calculate the rest-frame SED for each individual source by de-redshifting the observed MIRI band wavelengths, e.g., the F1500W band for a $z=2$ source is shifted to rest-frame 5~$\mu$m.
In this procedure, we only adopt MIRI photometry with $\rm S/N>3$ to guarantee good data quality.
We obtain each individual SED by linearly interpolating the MIRI flux density and normalizing each resulting SED at rest-frame 3.6~$\mu$m.
If a SED does not cover 3.6~$\mu$m, we discard this individual SED in the subsequent median-SED estimation.
\fst{The numbers of SEDs without 3.6~$\mu$m coverage are 263 (SF), 79 (mixed), 7 (AGN).}
The value of normalization wavelength (3.6~$\mu$m) is chosen to optimize the AGN sample size (our main focus in this work) as well as to avoid strong emission features (PAHs and lines).
These individual SEDs are displayed as the grey curves in Fig.~\ref{fig:composite_sed}.
Finally, we obtain the median SED at each wavelength by taking the median value of all individual SEDs that cover this wavelength.
We estimate the associated \fst{1$\sigma$ scatter using the 16\%--84\% percentiles}. 
In this step, if the number of individual SEDs drops below 5, we do not calculate the median SED, and this criterion leads to the shortest and longest wavelengths of the median SED (differing by source types).
Note that we do not divide our sources into different redshift bins and study the median SEDs for each bin, mainly because our sample sizes (especially for AGN) are limited (see \S\ref{sec:caveat}).
Also, we use the interpolated SEDs directly instead of the best-fit SEDs from \textsc{cigale}, because choosing the latter would be more strongly model dependent. 

Fig.~\ref{fig:composite_sed} displays the resulting median SED for each type. 
On the left panel (SF), we also over-plot the observed SEDs of two local star-forming galaxies, NGC~5992 and NGC~6090, from \cite{brown14}, and these two SEDs appear similar to each other. 
The broad shape of our median SF SED is generally similar to these two local-SF SEDs: all three SEDs are relatively flat at wavelengths below $\sim 5\ \mu$m and rise toward longer wavelengths due to the PDR emission. 
However, there are also some differences. 
The PAH emission features appear to be significantly weaker in our median SED.  
This difference might be mainly caused by the broad-band nature of our MIRI photometry ($\lambda/\Delta \lambda \sim 5$), i.e., we lack the spectral resolution to highlight the intrinsic PAH intensities.  
Also, the photo-$z$ uncertainties inevitably weaken the observed PAH strength in the median SED. 
Detailed quantification of the PAH strength for our MIRI sources can be performed with detailed SED modelling, but this is beyond the scope of this work. 
Future MIRI medium-resolution spectrometer (MRS) programs targeting distant SF galaxies will allow a robust comparison of the PAH strength versus local galaxies \fst{\citep[approved MRS programs, e.g.,][]{colina_robledo17, pope21}}. 
At wavelengths of $\lambda \sim 2\ \mu$m, the local-SF SEDs have higher fluxes than our median SED, indicating that the former have stronger stellar emission peaking at near-IR wavelengths. 
This is understandable, because our sources are MIRI-selected distant SF galaxies and they likely have relatively stronger mid-IR PDR emission compared to the near-IR stellar emission. 

The right panel in Fig.~\ref{fig:composite_sed} compares our median AGN SED versus the median SED of hot dust-obscured galaxies (hot DOGs; \citealt{fan16}) and the average SED of Seyfert~2 galaxies \citep{videla13}.
Hot DOGs are infrared-luminous ($L_{\rm IR} \gtrsim 10^{13}\ L_\odot$) galaxies  selected by \textit{Wide-field Infrared Survey Explorer} (\textit{WISE}) typically at $z\gtrsim 2$, and they are considered to be powered by fast-growing heavily obscured AGN \citep[e.g.,][]{eisenhardt12, wu12}. 
Seyfert~2 galaxies are type~2 obscured AGN in the local universe, and the average SEDs are based on photometry from the nuclear regions to avoid starlight from the outskirt regions. 
\citet{videla13} published average Seyfert~2 SEDs from two populations: one with near-IR excess (i.e., potential stellar emission) and the other without. 
We choose the latter to focus on the AGN emission. 
From Fig.~\ref{fig:composite_sed}, our median AGN SED is similar to the hot DOG and Seyfert~2 SEDs.
This similarity indicates that the MIRI-selected AGN have comparable physical dust-obscuration structures versus hot DOGs and Seyfert~2s, at least in the ensemble sense. 
We stress that our MIRI-selected AGN are likely more representative in the BH growth history, because hot DOGs are too rare with a very low number density (a few in 100~deg$^2$; e.g., \citealt{assef15}) and \fst{nearby Seyfert~2s are systems in which} BHs have largely finished their mass growth \citep[e.g.,][]{yang18}. 

\begin{figure}
    \centering
    \includegraphics[width=\columnwidth]{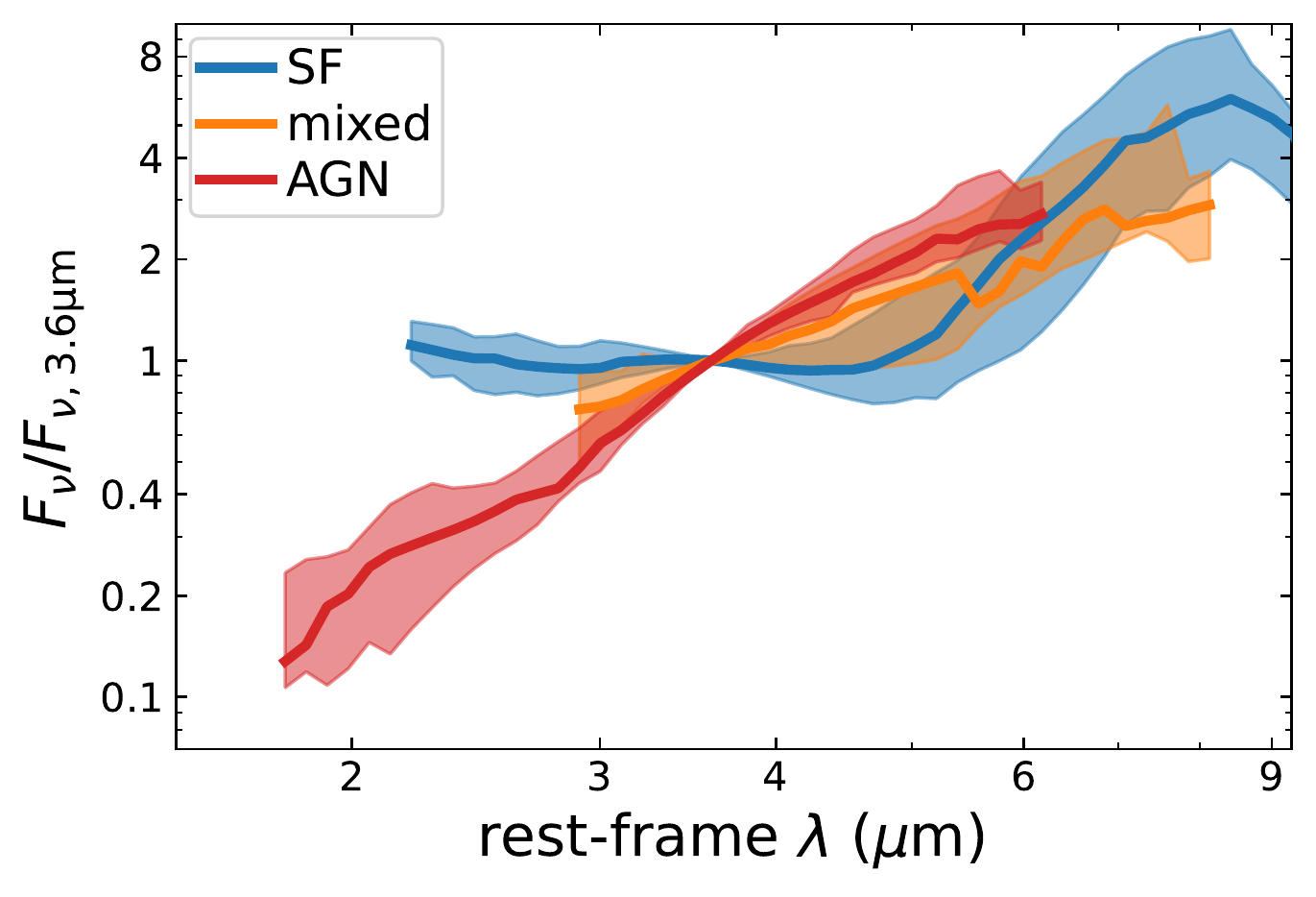}
    \caption{Comparison between different types of rest-frame median SEDs (same as in Fig.~\ref{fig:composite_sed}). The blue, orange, and red represent SF, mixed, and AGN types, respectively. 
    The shaded regions indicate the corresponding $1\sigma$ \fst{scatters}, estimated as 16\%--84\% percentiles.
    The AGN SED is distinctive from the SF SED. 
    These median SEDs and their associated errors are publicly available together with the online version of this paper.
    }
    \label{fig:sed_pk}
\end{figure}

Fig.~\ref{fig:sed_pk} compares the median SEDs from different source types. 
The SF and AGN SEDs have distinctive shapes. 
The SF SED is significantly flatter than the AGN SED at $z \lesssim 5$, because the former has relatively strong stellar emission peaking at near-IR wavelengths. 
In fact, the IRAC-based color-color AGN selection methods are largely based on this difference in SF versus AGN SED shapes \citep[e.g.,][]{stern05, lacy07, donley12}.  
As expected, the mixed SED appears to fall between the SF and AGN SEDs, but it tends to be closer to the latter. 
We publicly release the median SEDs and their associated \fst{scatters} (as displayed in Fig.\ref{fig:sed_pk}), in the format of electronic tables along with the online version of this paper.

\fst{Finally, we point out that the median SEDs are not completely model-independent, although they are based on the interpolated photometry instead of the best-fit models.
The source classification relies on model fitting, and thus the resulting median SEDs (especially the broad shapes) mimics the corresponding models by construction.
The PAH features might be less dependent on the SED-based selections, compared to the broad SED shapes.
}

\subsection{Black-hole growth history}
\label{sec:bhad}
X-ray AGN selections have several advantages of, e.g., high purity and good sensitivity (see \S\ref{sec:intro}). 
Thanks to these benefits, X-ray surveys have been widely employed to study the BH cosmic accretion history, often characterized by black-hole accretion density (BHAD), i.e., total BH growth rate per comoving volume.
However, X-ray photons can be heavily obscured if the obscuring materials are abundant (\S\ref{sec:intro}). 
This leads to the suspicion that even the deepest X-ray surveys so far might suffer from potential incompleteness issues when characterising the BHAD evolution (e.g., \citealt{hickox18}; 
\fst{\citealt{carroll21}}; \citealt{yang21b}).
It is thereby crucial to complement the X-ray results with BHAD measurements based on other wavelengths. 
In this section, we estimate BHAD based on our SED-fit results and compare it with X-ray measurements from the literature.

\begin{figure}
    \centering
    \includegraphics[width=\columnwidth]{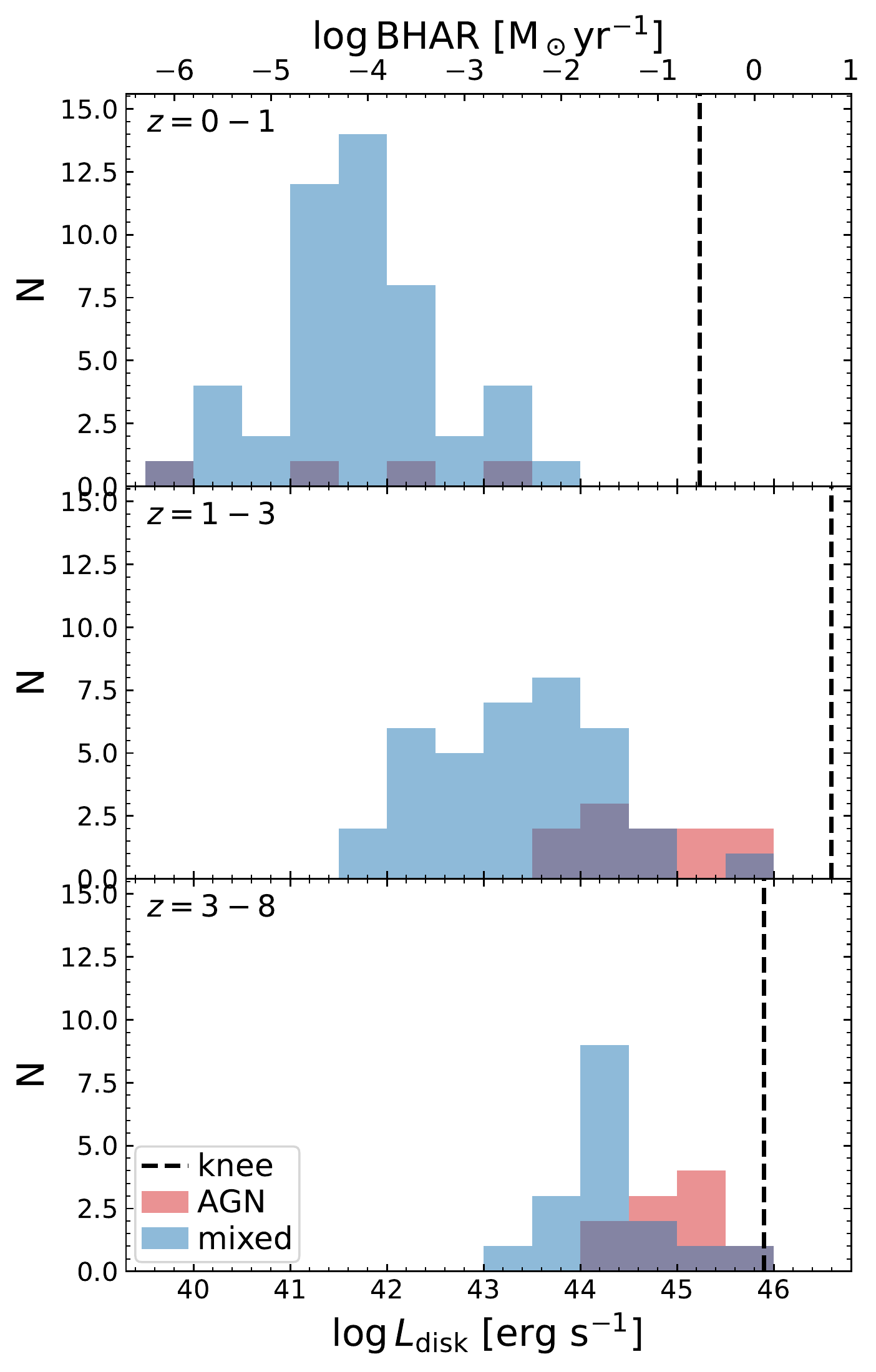}
    \caption{$L_{\rm disk}$ distributions for AGN (red, with $\fracA \geq 0.5$) and mixed (blue, with $0.1 < \fracA < 0.5$) objects.
    Different panels are from different redshift bins as labeled. 
    \fst{The vertical dashed lines indicate the typical knee luminosities from \cite{shen20}.}
    The corresponding BHAR is also marked at the upper axis. 
    $L_{\rm disk}$ is numerically equivalent to angle-averaged $L_{\rm bol}$ (see \S\ref{sec:bhad}).
    }
    \label{fig:Ldisk_hist}
\end{figure}

To estimate BHAD, we first calculate the BH accretion rate (BHAR) for each source classified as AGN or mixed using 
\begin{equation}
\begin{split}
    {\rm BHAR} &= \frac{L_{\rm disk} (1-\epsilon)}{\epsilon c^2} \\
    &= \frac{1.59 L_{\rm disk}}{10^{46}
            {\rm\ erg\ s^{-1}}} 
            {\rm\ M_\odot\ yr^{-1}}
\end{split}
\end{equation}
where $c$ is the speed of light and $\epsilon$ is the \fst{radiative} efficiency.
We assume a conventional value of $\epsilon=0.1$ here \citep[e.g.,][]{brandt15}. 
$L_{\rm disk}$ is the viewing angle-averaged intrinsic accretion-disk luminosity from our SED fits (i.e., the ``accretion\_power'' quantity in the \textsc{cigale} output; see \citealt{yang18}).
It is numerically equivalent to the angle-averaged AGN bolometric luminosity due to energy conservation. 
The $L_{\rm disk}$ (BHAR) distributions at different redshifts are displayed in Fig.~\ref{fig:Ldisk_hist}.

We then add up all the BHAR in a given redshift bin, and divide the total BHAR by the comoving volume in the redshift bin sampled by the survey area (a total of 9.2~arcmin$^2$ for the 4 pointings).
This yields our BHAD estimate for a given redshift bin, i.e., 
\begin{equation}
    {\rm BHAD} = \frac{\sum_{i} {\rm BHAR}_i}{V_c},
\end{equation}
where ${\rm BHAR}_i$ is the BH accretion rate for the $i$-th source (AGN or mixed type; see below) in the redshift bin; $V_c$ is the comoving volume for the redshift bin sampled by the MIRI area (9.2~arcmin$^2$), and the numerical value is calculated using {\sc astropy.cosmology}. 
Note that we do not correct for incompleteness here.
The correction would require detailed knowledge of our MIRI-selection sensitivity as well as AGN IR luminosity function, and that is beyond the scope of this work. 
We qualitatively discuss the effects of incompleteness in \S\ref{sec:caveat}. 

\begin{figure}
    \centering
    \includegraphics[width=1\columnwidth]{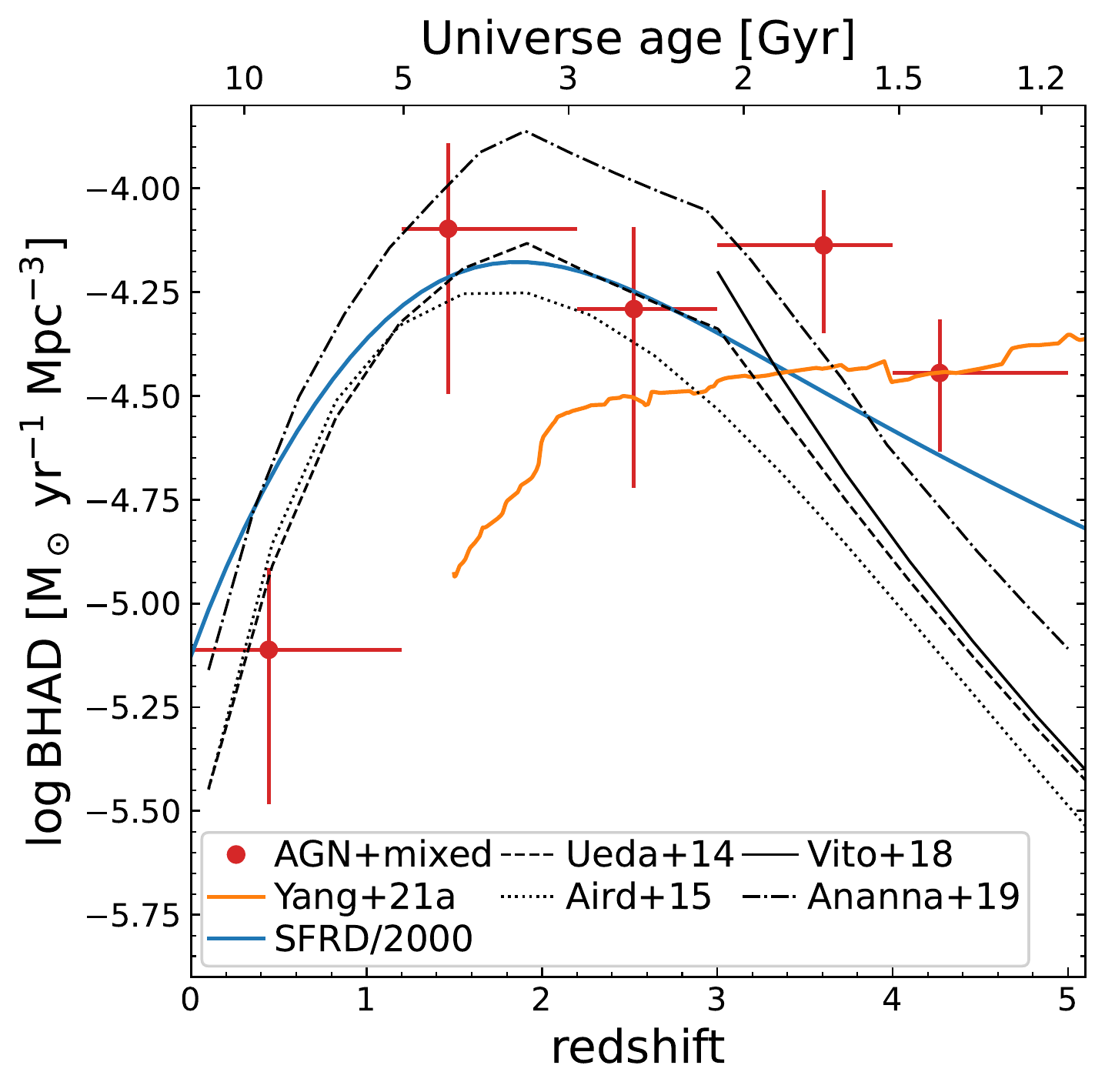}
    \caption{Black hole accretion density as a function of redshift. 
    The red data points are derived based on our MIRI-selected AGN; the orange ones are from AGN$+$mixed sources. 
    The vertical error bars \fst{consider both bootstrap and SED-fit (Bayesian) uncertainties}; the horizontal error bars indicate the redshift bins.
    The orange curve representing the inferred BHAR for bulge-dominated galaxies is from \cite{yang21b}, and is thereby a lower bound on the total accretion rate.
    The black curves are from \hbox{X-ray} measurements with references marked in the legend (see \citealt{yang21b} for details).
    At high redshifts of $z\gtrsim 3$, our MIRI-based BHAD values are significantly higher than the corresponding \hbox{X-ray} measurements, but are consistent with the \cite{yang21b} lower bound.
    The blue curve represents SFRD \citep{madau14} down-scaled by a factor of 2000, to roughly match the X-ray-based BHAD peak at $z\sim 2$. 
    }
    \label{fig:bhad}
\end{figure}

The resulting BHAD versus redshift is displayed in Fig.~\ref{fig:bhad} as red data points. 
The redshift bins (same for the red and orange points) are indicated by the horizontal error bars.
We estimate the uncertainties with a bootstrap method, using \textsc{astropy.stats.bootstrap} \citep{astropy}.
The bootstrap resample number is 1000, and we adopt the 16--84\% percentile as 1$\sigma$ confidence limit for each data point.
\fst{The bootstrap error ($\delta \rm BHAD_{boot}$) only accounts for statistical fluctuations due to limited sample sizes.
To also consider the errors in our SED fits, we propagate the $L_{\rm disk}$ Bayesian errors (from the \textsc{cigale} output) into the BHAD error ($\delta \rm BHAD_{Bayes}$), using standard error propagation.
Finally, We estimate the final BHAD uncertainty as $\sqrt{ (\delta \rm BHAD_{boot})^2 +  (\delta BHAD_{Bayes})^2 }$.  
$\delta \rm BHAD_{Bayes}$ is smaller than or comparable to the corresponding $\delta \rm BHAD_{boot}$ and their ratios are 0.2--1.0. 
}

In Fig.~\ref{fig:bhad}, we also show some BHAD curves from the literature. 
The \cite{yang21b} curve is based on the observed SFH of bulge-dominated galaxies at $z=0.7$--1.5 with the assumption of the bulge BHAR-SFR relation \citep{yang19}, and thus this BHAD measurement is an indirect method.
The \cite{yang21b} BHAD should be considered as a lower limit, as it only accounts for the BH growth within bulge-dominated galaxies. 
The other literature BHAD curves are based on X-ray data \citep{ueda14, aird15, vito18, ananna19}, each including a wide range of surveys (from wide-shallow to narrow-deep) to completely sample the entire X-ray AGN population. 
The X-ray measurements are compiled by \cite{yang21b}, assuming the same radiation efficiency ($\epsilon = 0.1$) as our MIRI-based BHAD.

From Fig.~\ref{fig:bhad}, our MIRI-based BHAD roughly agrees with the X-ray results at low redshifts of $z\lesssim 3$. 
However, the X-ray BHAD curves drop sharply toward higher redshifts.
In contrast, our BHAD does not have significant evolution at $z\approx 1$--5. 
This difference in evolution leads to the fact that our BHAD is much higher than the X-ray values by $\sim 0.5$~dex at $z \gtrsim 3$. 

The results above indicate that MIRI detects a great number of AGN that are missed by X-ray observations. 
Indeed, none of our $z>3$ AGN/mixed objects are detected by the \textit{Chandra}\ X-ray observations \citep{nandra15}.
This lack of X-ray signals is likely due to heavy obscuration at high redshifts (see \S\ref{sec:intro}).  
In the early universe, the obscuring gas is likely more abundant and compact compared to low redshifts, resulting in a high column density that is capable of heavily damping the X-ray emission \citep[e.g.,][]{gilli22}.   
On the other hand, heavy AGN obscuration should be associated with high IR luminosities due to strong dust absorption and re-emission following the energy-conservation law (\S\ref{sec:intro}).
This explains why MIRI is able to detect the X-ray missed high-$z$ AGN.

\fst{Now, we quantitatively constrain the X-ray obscuration $N_{\rm H}$ via a stacking analysis.
We apply the same stacking method of \cite{yang19} to our AGN at $z=3$--5 (10 sources, all undetected in X-ray). 
In brief, this method extracts X-ray count rates at the $H$-band position of each source, and then convert the average count rate to the average $L_{\rm X}$ (X-ray luminosity in rest-frame 2--10~keV).
The details of the stacking method are described in Sec.~2.4 of \cite{yang19}.
For our high-$z$ AGN sample, the stacked $L_{\rm X}$ is consistent with zero (likely due to our limited sample size), having a $3\sigma$ upper limit of $2.7\times 10^{42}$~erg~s$^{-1}$.
On the hand, based on our fitted $L_{\rm disk}$ (Fig.~\ref{fig:Ldisk_hist}), we have estimated an average intrinsic $L_{\rm X}$ of $5.9\times 10^{43}$~erg~s$^{-1}$, assuming an X-ray bolometric correction factor of 22.4 \citep[e.g.,][]{vasudevan07}.
From this expected intrinsic $L_{\rm X}$ and the observed $3\sigma$ upper limit of $L_{\rm X}$, we estimate a damping factor of at least $\sim 22$ ($5.9\times 10^{43}$ divided by $2.7\times 10^{42}$) for the AGN X-ray radiation. 
To reach this level of obscuration, the $N_{\rm H}$ should be at least $\sim 2.0 \times 10^{24}$~cm$^{-2}$,\footnote{This is estimated with \textsc{pimms}, https://cxc.harvard.edu/toolkit/pimms.jsp} above the Compton-thick threshold $1.5 \times 10^{24}$~cm$^{-2}$.
Therefore, we conclude that our MIRI-detected high-$z$ AGN are heavily obscured in X-ray on average.
We have also applied the analysis above to the sample of AGN and mixed sources (25 objects) at $z=3$--5, and reach a similar result. 
}
\fst{We note that it is infeasible to directly infer X-ray $N_{\rm H}$ from the obscuration estimated from our SED fitting of the UV-to-IR photometry. 
This is because UV-to-IR radiation is dominantly reprocessed by dust, while X-ray is dominantly reprocessed by gas \citep[e.g.,][]{hickox18}. 
Actually, the \texttt{skirtor2016} model we used in SED fitting does not include any gas at all \citep{stalevski12, stalevski16}, and thus we are not able to infer $N_{\rm H}$ from our SED modelling.}

Fig.~\ref{fig:bhad} also displays the redshift evolution of SFRD from \cite{madau14}.
In the plot, the SFRD is numerically down-scaled by a factor of 2000, to roughly match the X-ray-based BHAD peak at $z \sim 2$. 
Qualitatively similar to the BHAD measurements from X-ray data, the SFRD also decreases at $z\gtrsim 2$. 
This behavior leads to that the SFRD falls below our BHAD at $z\gtrsim 3$, indicating the BHAD/SFRD ratio rises at high redshifts. 
This result poses a strong question to theorists, as current simulations predict that the BHAD/SFR ratio should drop (or at most remain constant) in the early universe \citep[e.g.,][]{habouzit21, zhang23}. 
New theoretical models are perhaps needed for more efficient BH growth at high redshifts. 

\subsection{Caveats}
\label{sec:caveat}
Our work represents one of the first papers using \textit{JWST}/MIRI to search for obscured AGN. 
Currently, we are still in the early stage of \textit{JWST} operation, and thereby this paper inevitably has some limitations, as we discuss below. 

A major limitation of this work is the sample size, only a total of 560 sources (433 SF, 102 mixed, and 25 AGN), over a wide redshift range of $z\approx 0$--5.
This relatively small sample size is due to the survey area we have, i.e., only four pointings with a total area of 9~arcmin$^2$.
The sample sizes (especially for AGN and mixed types) prevent us from studying the redshift evolution of the median SEDs (\S\ref{sec:comp_sed}). 

For the BHAD measurements (\S\ref{sec:bhad}), the contributions from rare luminous AGN (such as hot DOGs) are missed, because our MIRI survey area is not sufficiently large to capture them.
\fst{Indeed, most of our sources lie below the knee luminosity (dashed lines in Fig.~\ref{fig:Ldisk_hist}).}
For comparison, the least luminous hot DOGs have $L_{\rm disk} \sim 10^{47}\rm erg\ s^{-1}$, \fst{above the knee luminosity} \citep[e.g.,][]{fan16}.
In addition, our sample does not include unobscured type~1 AGN, which is also generally luminous, due to our limited survey area (see \S\ref{sec:sed}).
Therefore, the intrinsic BHAD might be higher than our estimated values.  
However, we note that this bias actually strengthens our main conclusion, i.e., the BH growth in the early universe is stronger than we thought based on X-ray detections. 
On the faint side, our MIRI data could miss low-luminosity AGN especially at high redshifts.
For example, all the sources at $z>3$ have $L_{\rm disk} \gtrsim 10^{43}\ \rm erg\ s^{-1}$ in Fig.~\ref{fig:Ldisk_hist}.
\fst{Another AGN population we could miss is those mainly obscured by galactic-scale dust \citep[e.g.,][]{gilli22}, of which the reprocessed emission is beyond the MIRI wavelength coverage (see \S\ref{sec:sed}).
Such biases again go to the direction of strengthening our main conclusion.}
\fst{From the discussion above, the contributions from many AGN could be missed in the MIRI-based BHAD measurements.
Therefore, our estimation effectively serves as a lower-limit constraint to the intrinsic complete BHAD.}

Another limitation of this work is the lack of secure spec-$z$ for most sources in our sample, especially the high-$z$ AGN/mixed objects which are crucial to our main results. 
Although the photo-$z$ quality appears high (\S\ref{sec:sed_res}), the spec-$z$ sources used in this quality assessment are all at low redshifts of $z<3$ (see Fig.~\ref{fig:z_vs_z}).  
To have a rough evaluation of the quality at high redshifts, we compare the photo-$z$ of our $z>3$ sources versus the CANDELS-EGS photo-$z$ \citep{stefanon17}.
Only 1 (out of 36) sources have fractional differences $>15\%$, indicating that our selected high-$z$ sample is also supported by the CANDELS-EGS work. 
\fst{The difference between our $z>3$ BHAD and the X-ray measurements are $\sim 0.5$~dex (Fig.~\ref{fig:bhad}).
Therefore, if this large difference were due to photo-$z$ errors, most ($\gtrsim 70\%$) of our photo-$z$ estimations for our high-$z$ sources would have to be spurious. 
}

Finally, our SED-based method is model dependent. 
It is possible that our adopted IR models (see \S\ref{sec:sed}) are not suitable especially for MIRI-selected high-$z$ objects, because these sources are at a low mid-IR flux level that has not been well studied. 
For example, if the 3.3~$\mu$m PAH feature is exceptionally strong, the observed MIRI SED of a $z \approx 4$--5 object could be very red and thereby misclassified as an AGN \citep[e.g.,][]{magnelli08}. 
This is because the rest-frame 3.3~$\mu$m is shifted to F1800W or F2100W (our reddest MIRI bands).
To address this potential issue, we re-perform our analyses but discarding the F1800W and F2100W data.
The resulting BHAD is consistent with the value based on all MIRI bands, still significantly higher than the X-ray measurements at $z\approx 4$--5.
Therefore, the possible model uncertainties about 3.3~$\mu$m PAH are unlikely to affect our main conclusion qualitatively. 

\section{Summary and Future Prospects}
\label{sec:sum}
In this work, we have performed SED analysis for a MIRI-detected sample in the CEERS survey. 
Our main results are summarized below. 

\begin{itemize}

    \item Our analyses are based on the four MIRI red pointings of the CEERS survey ($\sim 9\ \rm arcmin^2$). 
    From the CEERS MIRI catalog, we have selected sources with at least two MIRI-band $\rm S/N>3$, and compiled other multiwavelength (UV-to-IRAC~4) photometry from the CANDELS catalog. 
    We have a total of 560 MIRI-detected sources.
    
    \item We have performed SED modelling for the 560 sources with \textsc{cigale}.
    In this procedure, we have fitted the redshift and other source properties (e.g., stellar $e$-folding time and age) simultaneously. 
    We compared our results to the available spec-$z$ of 29 objects in the CANDELS-EGS catalog. 
    We found  $\nmad=0.032$ and no outliers with $>15\%$ uncertainties, indicating our photo-$z$ estimate is robust. 
    Based on the best-fit $\fracA$ (fractional AGN luminosity at rest-frame 3--30~$\mu$m), we have classified each source into SF, mixed, and AGN, respectively. 

    \item We have derived median rest-frame SEDs for the three types using MIRI photometry.  
    The median SF SED has a similar broad shape compared to local SF galaxies, but the PAH features appear to be weaker, probably due to the broad wavelength coverage of the MIRI filters and/or redshift uncertainties. 
    The median AGN SED is very close to the typical SEDs of Hot DOGs and Seyfert~2s, suggesting they are intrinsically the same type of objects, i.e., actively accreting but obscured black holes. 
    We publicly release our median SEDs along with this paper.  

    \item We have studied cosmic BH accretion history by estimating the BHAD as a function of redshift based on our classified AGN (and mixed) type of sources.  
    The result agrees with the X-ray measurements at $z \lesssim 3$.
    However, our MIRI-based BHAD becomes significantly higher than the latter toward higher redshifts, and the difference is $\sim 0.5$~dex at $z \gtrsim 3$.
    We interpret this difference as that MIRI is able to detect many heavily obscured AGN in the early universe. 
    
\end{itemize}

With the accumulation of total \textit{JWST} observing time, there will be more and more archival MIRI imaging data suitable for such studies as this work.
This will effectively address the potential issues related to the small sample size, and also allows studying the redshift evolution of the median SEDs (\S\ref{sec:caveat}).  
Future MIRI/MRS follow-up observations of the imaging-selected SF/AGN targets will also be useful:
MRS can reveal the redshift evolution of PAH strength among SF galaxies (\S\ref{sec:comp_sed}); 
MRS can provide secure spectroscopic redshifts for our high-$z$ AGN candidates, essentially testing our conclusion of the BHAD excess compared to X-ray measurements (\S\ref{sec:bhad}).

An alternative approach to probe AGN at high redshifts is mining the existing \textit{JWST} spectroscopic database (e.g., NIRSpec Multi-Object Spectroscopy and NIRCam grism slitless spectroscopy), many of which target high-$z$ galaxy candidates. 
Based on the spectra, the presence of AGN can be identified by searching for broad emission lines and/or narrow high-ionization lines \citep[e.g.,][]{cleri22, kocevski23}.
The spectroscopic selections can potentially probe AGN at the highest redshifts.
For example, the recent work of \cite{larson23} reported a broad-line AGN at $z=8.679$, the earliest BH discovered so far.  

\begin{acknowledgements}
We thank the referee for helpful feedback that improved this work.
We acknowledge the hard work of our colleagues in the CEERS collaboration and everyone involved in the \textit{JWST} mission.
GY, KIC and EI acknowledge funding from the Netherlands Research School for Astronomy (NOVA). 
KIC and VK acknowledge funding from the Dutch Research Council (NWO) through the award of the Vici Grant VI.C.212.036.  
CP thanks Marsha and Ralph Schilling for generous support of this research. 
TAH is supported by an appointment to the NASA Postdoctoral Program (NPP) at NASA Goddard Space Flight Center, administered by Oak Ridge Associated Universities under contract with NASA. 
This work acknowledges support from the NASA/ESA/CSA James Webb Space Telescope through the Space Telescope Science Institute, which is operated by
the Association of Universities for Research in Astronomy, Incorporated, under NASA contract NAS5-03127. Support for program No. JWST-ERS01345 was provided through a grant from the STScI under NASA contract NAS5-03127.
Some of the data presented in this paper were obtained from the Mikulski Archive for Space Telescopes (MAST) at the Space Telescope Science Institute.The specific observations analyzed can be accessed via \dataset[http://dx.doi.org/10.17909/agda-2w34]{http://dx.doi.org/10.17909/agda-2w34}. STScI is operated by the Association of Universities for Research in Astronomy, Inc., under NASA contract NAS5–26555. Support to MAST for these data is provided by the NASA Office of Space Science via grant NAG5–7584 and by other grants and contracts.
\end{acknowledgements}

\software{
{\sc astropy} \citep{astropy},
{\sc cigale} \citep{boquien19, yang20, yang22}
}

\bibliography{all}{}
\bibliographystyle{aasjournal}

\end{CJK*}
\end{document}